\documentstyle[12pt,epsfig]{article}
\newcommand{\beq}{\begin{equation}}
\newcommand{\eeq}{\end{equation}}
\newcommand{\bea}{\begin{eqnarray}}
\newcommand{\eea}{\end{eqnarray}}
\newcommand{\gsim}{\raisebox{-0.07cm}{$\:\stackrel{>}{{\scriptstyle
 \sim}}\: $} }

\newcommand\MSb{$\overline{\mbox{MS}}$}
\begin{document}
\setlength{\parskip}{0.25cm}
\setlength{\baselineskip}{0.55cm}
\begin{titlepage}

\noindent
{\tt hep-ph/0007362} \hfill INLO-PUB 09/00 \\ 
\hspace*{\fill} August 2000 \\
\vspace{1.5cm}
\begin{center}
\Large
{\bf Improved Approximations for the} \\
\vspace{0.15cm}
{\bf Three-Loop Splitting Functions in QCD} \\
\vspace{3.0cm}
\large
W.L. van Neerven and A. Vogt \\
\vspace{0.8cm}
\normalsize
{\it Instituut-Lorentz, University of Leiden \\
\vspace{0.1cm}
P.O. Box 9506, 2300 RA Leiden, The Netherlands} \\
\vspace{6.0cm}
{\bf Abstract}
\vspace{-0.3cm}
\end{center}
We update our approximate parametrizations of the three-loop splitting
functions for the evolution of unpolarized parton densities in 
perturbative QCD. The new information taken into account is given by
the additional Mellin moments recently calculated by Retey and 
Vermaseren. The inclusion of these constraints reduces the 
uncertainties of our approximations considerably and extends their 
region of applicability by about one order of magnitude to lower 
momentum fractions~$x$.

\vspace{0.8cm}
\begin{center}
\end{center}

\end{titlepage}
 
\noindent
In order to achieve a high accuracy of the predictions of perturbative
QCD for hard processes, the calculations need to transcend the standard 
next-to-leading order (NLO) approximation. For processes with initial-%
state hadrons, the next-to-next-to-leading order (NNLO) expressions
include the three-loop splitting functions. The computation of these 
functions is under way \cite{MV1}, but will not be completed in the 
near future \cite{RV}.

Partial results have already been obtained [3--9], however, most 
notably the five lowest even-integer moments for the flavour 
non-singlet combination entering electromagnetic deep-inelastic 
scattering (DIS) \cite{spfm1}, and four moments for the singlet 
splitting functions~\cite{spfm2}. 
In refs.~\cite{NV1,NV2} we have demonstrated that this information --- 
due to the smoothening effect of the ubiquitous convolution with the 
initial parton densities and the small size of the corrections --- is 
fully sufficient for momentum fractions $x\,\gsim\, 0.1$ and leaves 
only small uncertainties down to $x \simeq 10^{-3}$ at scales above 
about 10 GeV$^2$. We have provided approximate parametrizations for the 
three-loop splitting functions, including quantitative estimates of 
their residual uncertainties. These results have already been applied 
\cite{MRST} to structure functions in DIS and Drell-Yan cross sections 
at hadron colliders, for which the subprocess cross sections have been 
computed up to NNLO \cite{ZvN,c2DY}.

Very recently the fixed-moment calculations of refs.~\cite{spfm1,spfm2}
have been extended, using improved computing resources, up to the 
twelfth moment \cite{RV}. For the first time also (odd) moments of the 
three-loop valence splitting functions have been obtained there. These 
results provide a severe check of our approximation procedure, as the 
latter led to rather tight predictions for the (tenth and) twelfth 
moments. This test is passed by the results of refs.~\cite{NV1,NV2}.
In this letter we update these parametrizations by including the 
moments of ref.~\cite{RV} in the derivation. As a result the residual
uncertainties are greatly reduced, and the region of safe applicability 
is extended by about one order of magnitude in $x$, an improvement most 
relevant for applications to structure functions at HERA \cite{MRST}.

Our notations for the parton densities and splitting functions are as 
follows: the non-singlet combinations of quark and antiquark densities, 
$q_i$ and $\bar{q}_i$, are given by
\beq
\label{qns}
  q_{{\rm NS}}^{\pm} \: = \: q_i \pm \bar{q}_i - (q_k \pm \bar{q}_k)
  \:\: , \quad
  q_{\rm NS}^V \: = \:\sum_{r=1}^{N_f} (q_r - \bar{q}_r) \:\: .
\eeq
$N_f$ stands for the number of effectively massless flavours.
The corresponding splitting functions are denoted by $P_{\rm NS}^{\pm}$
and $P_{\rm NS}^V \! \equiv \! P_{\rm NS}^- + P_{\rm NS}^{S}$. The 
latter function, $P_{\rm NS}^{S}$, occurs for the first time at 
$O(\alpha_s^3)$. Except for the vanishing of the first moment, it was 
unknown before the calculation of ref.~\cite{RV}. The evolution 
equations in the singlet sector are written as
\beq
\label{sgev}
  \frac{d}{d \ln \mu_f^2}
  \left( \begin{array}{c} \!\Sigma\! \\ g  \end{array} \right)
  \: = \: \left( \begin{array}{cc} P_{\rm NS}^+ + P_{\rm PS}^{\,} 
  & P_{qg} \\ P_{gq} & P_{gg} \end{array} \right) \otimes
  \left( \begin{array}{c} \!\Sigma\! \\ g  \end{array} \right) 
  \:\: , \quad
  \Sigma \: = \: \sum_{r=1}^{N_f} ( q_r + \bar{q}_r ) \:\: .
\eeq
Here $g$ represents the gluon density, and $\otimes$ denotes the Mellin 
convolution. The expansion of all these splitting functions in powers
of the running coupling constant $\alpha_s$ reads
\beq 
\label{pexp}
   P(x,\alpha_s) \: = \: a_s \, P^{(0)}(x) + a_s^2 \, P^{(1)}(x) 
   + a_s^3 \, P^{(2)}(x) + \ldots \:\: \mbox{with} \quad 
   a_s \:\equiv\: \alpha_s/4\pi \:\: , 
\eeq
if the renormalization and factorization scales are identified, 
$\mu_r = \mu_f$. The additional terms for $\mu_r \neq \mu_f$ are 
exactly known up to NNLO and need not to be considered here.
  
Let us briefly illustrate our approximation procedure for the case of 
the $N_f$ term, $P_{qg,1}^{(2)}(x)$, of the gluon-quark splitting 
function $P_{qg}$ which dominates the small-$x$ evolution of the quark 
densities. Taking into account the small-$x$ result of ref.\ 
\cite{CH94}, the expected form of this function in the \MSb\ scheme 
employed throughout this paper is given by
\beq
\label{pqg1}
  P_{qg,1}^{(2)}(x) \: = \: \sum_{m=1}^{4} A_m \ln^m (1-x) 
  + f_{\rm smooth}(x) + \sum_{n=1}^{4} B_n \ln^n x + \frac{C}{x}
  - \frac{896}{27}\, \frac{\ln x}{x} \:\: ,
\eeq
where $f_{\rm smooth}$ is finite for $ 0 \leq x \leq 1$. We choose
three or two of the large-$x$ logarithms, one or two smooth functions
(mainly low powers or simple polynomials of $x$) and two of the 
small-$x$ terms ($x^{-1}$ and $\ln x$ or $\ln^2 x$). Their coefficients 
are then determined from the known six moments \cite{RV,spfm2}. Varying 
these choices we arrive at the about 50 approximations compared in 
Fig.~1. The two representatives spanning the error band for most of the 
$x$-range are finally selected as our estimates for $P_{qg,1}^{(2)}$ 
and its residual uncertainty.

Analogous procedures are applied to the $N_f^0$ and $N_f^1$ terms of
all other functions $P^{(2)}(x)$. The non-singlet $N_f^2$ contribution
is known \cite{Gra1}. The singlet $N_f^2$ pieces are smaller in 
absolute size and uncertainty than the $N_f^0$ and $N_f^1$ terms, hence 
for them it suffices to select just one central representative.
Note that the previous information, four (five) moments in the singlet 
($P^{+}_{\rm NS}$) sector, respectively, was not sufficient to fix
the coefficients of the subleading small-$x$ terms $\propto x^{-1}$ 
($\ln^3 x$) from the moments. Thus we had to resort to conservatively
varied, educated guesses inspired by the NLO results. Except for the 
case of $P^{(2)+}$ (where the rightmost pole in Mellin space is not 
one, but two units away from the lowest calculated moment) we can now 
dispense with these ad hoc estimates.

We now write down our improved parametrizations, using the 
abbreviations
\beq
  L_0 \:\equiv\: \ln x \:\: , \quad L_1 \:\equiv\: \ln (1-x) \:\: .
\eeq
Two approximations, denoted by $P^{(2)}_A$ and $P^{(2)}_B$, are 
provided for each function. Where both are present, the $N_f^0$ and 
$N_f^1$ terms have been combined such that the error bands are 
maximized at small $x$. The averages $1/2 \: [A+B]$ represent the
central results.

Our new expressions for the non-singlet splitting functions $P^{(2)\pm}
_{\rm NS}$ read
\bea
  P^{(2)-}_{{\rm NS}, A}(x)\! &=&
   1185.229\:(1-x)_+^{-1} + 1365.458\: \delta (1-x) - 157.387\: L_1^2 
   - 2741.42\: x^2 
  \nonumber \\ & & \mbox{} 
   - 490.43\: (1-x) + 67.00\: L_0^2 + 10.005\: L_0^3 + 1.432\:  L_0^4
   \nonumber \\ & & \mbox{} \hspace{-13mm}
   + N_f\: \{ - 184.765\: (1-x)_+^{-1} - 184.289\: \delta (1-x)
   + 17.989\: L_1^2 + 355.636\: x^2 
  \nonumber \\ & & \mbox{} 
   - 73.407\: (1-x)L_1 + 11.491\: L_0^2 + 1.928\:  L_0^3 \} \:\: + \:\:
   P^{(2)}_{{\rm NS},2}(x) 
  \nonumber \\
\label{nsm}
  P^{(2)-}_{{\rm NS}, B}(x)\! &=&
   1174.348\: (1-x)_+^{-1} + 1286.799\: \delta (1-x) + 115.099\: L_1^2 
   + 1581.05\: L_1  
  \nonumber \\ & & \mbox{}
   + 267.33\:  (1-x) - 127.65\: L_0^2 - 25.22\: L_0^3 + 1.432\: L_0^4
  \nonumber \\ & & \mbox{} \hspace{-13mm}
   + N_f\: \{
   - 183.718\: (1-x)_+^{-1} - 177.762\: \delta (1-x)
   + 11.999\: L_1^2 + 397.546\: x^2 
  \nonumber \\ & & \mbox{} 
   + 41.949\: (1-x) - 1.477\: L_0^2 - 0.538\:  L_0^3 \} \:\: + \:\:
  P^{(2)}_{{\rm NS},2}(x) 
\eea
and
\bea
  P^{(2)+}_{{\rm NS}, A}(x)\! &=&
   1183.762\: (1-x)_+^{-1} + 1347.032\: \delta (1-x) + 1047.590\: L_1 
   - 843.884\: x^2
  \nonumber \\ & & \mbox{}
   - 98.65\: (1-x) - 33.71\: L_0^2 + 1.580\: (L_0^4 + 4L_0^3)
  \nonumber \\ & & \mbox{} \hspace{-13mm}
   + N_f\: \{ - 183.148\: (1-x)_+^{-1} - 174.402\: \delta (1-x)
   + 9.649\: L_1^2 + 406.171\: x^2 
  \nonumber \\ & & \mbox{} 
   + 32.218\: (1-x) + 5.976\: L_0^2 + 1.60\:  L_0^3 \} \:\: + \:\:
   P^{(2)}_{{\rm NS},2}(x) 
  \nonumber \\
\label{nsp}
  P^{(2)+}_{{\rm NS}, B}(x)\! &=&
   1182.774\: (1-x)_+^{-1} + 1351.088\: \delta (1-x) - 147.692\: L_1^2 
   - 2602.738\: x^2   
  \nonumber \\ & & \mbox{}
   - 170.11 + 148.47\: L_0 + 1.580\: (L_0^4 - 4\, L_0^3)
  \nonumber \\ & & \mbox{} \hspace{-13mm}
   + N_f\: \{ - 183.931\: (1-x)_+^{-1} - 178.208\: \delta (1-x)
   - 89.941\: L_1 + 218.482\: x^2 
  \nonumber \\ & & \mbox{} 
   + 9.623  + 0.910\: L_0^2 - 1.60\:  L_0^3 \} \:\: + \:\:
  P^{(2)}_{{\rm NS},2}(x) \:\: .
\eea
The $\ln^4 x$ terms in Eqs.~(\ref{nsm}) and (\ref{nsp}) stem from 
ref.~\cite{BVns}. The exactly known $N_f^2$ contribution 
$P^{(2)}_{{\rm NS}, 2}(x)$ for both cases \cite{Gra1} is given in 
Eq.~(4.13) of ref.~\cite{NV1}. 
$P_{\rm NS}^{(2)+}(x)$ and $P_{\rm NS}^{(2)-}(x)$ are compared at $x<1$ 
in Fig.~2, for $N_f = 4$, to our previous approximations based on one 
moment less for $P_{\rm NS}^{(2)+}$ and mostly indirect information on 
$P_{\rm NS}^{(2)-}(x)$. Our present results are consistent with, but 
considerably more accurate than those of ref.~\cite{NV1}. 

In contrast to $P_{qg}$ and $P_{gq}$, the transition from one to two 
loops leads only to $\ln^1 (1-x)$ terms in $P_{\rm NS}$ and $P_{gg}$. 
Assuming correspondingly that no large-$x$ logarithms beyond 
$\ln^2 (1-x)$ occur in $P^{(2)\pm}_{\rm NS}$,
the coefficients of the $1/(1-x)_+$ term of the (also about 50) test 
functions considered for $P^{(2)-}_{\rm NS}(x)$ cover the range 
$\, 1167.3\: \ldots \: 1190.3\,$ for the $N_f^0$ part, and 
$\, -184.8\: \ldots$ $ -183.1\,$ for the $N_f^1$ contribution. The 
findings for $P^{(2)+}_{\rm NS}$, where one moment less is known, are 
consistent with these results. Note that terms $[\ln^k (1-x)/(1-x)]_+$ 
with $k \geq 1$ do not occur in the \MSb\ splitting functions, as 
proven in ref.~\cite{Ko89}.

The difference $P^{S}_{\rm NS}= P^V_{\rm NS}- P^-_{\rm NS}$ and the 
pure-singlet splitting function $P^{\,}_{\rm PS}$ in Eq.~(\ref{sgev}) 
result as $\, N_f(P^{S}_{q_i q_k} - P^{S}_{q_i \bar{q}_k})\,$ and 
$\, N_f(P^{S}_{q_i q_k} + P^{S}_{q_i \bar{q}_k})\,$, respectively, from 
the flavour independent (`sea') parts of the quark-quark and quark-%
antiquark splitting functions. The former combination carries the 
colour factor $d^{abc}d^{abc}$ which does not occur in $P^-_{\rm NS\,}$;
this fact facilitates the separations of the two terms in the results 
of ref.~\cite{RV}. Both $P^{S}_{\rm NS}(x)$ and $P^{\,}_{\rm PS}(x)$ 
vanish at $x = 1$, but are large at small $x$. The parametrizations
selected for the $a_s^3$ contributions to these two functions are given 
by 
\bea
  P^{(2)S}_{{\rm NS}, A}(x)\! &=&
   N_f\: \{ (1-x) ( - 1441.57\: x^2 + 12603.59\: x - 15450.01) 
   + 7876.93\: xL_0^2 
  \nonumber \\ & & \mbox{} 
   - 4260.29\: L_0 - 229.27\: L_0^2 + 4.4075\: L_0^3 \}
  \nonumber\\
\label{nss}
  P^{(2)S}_{{\rm NS}, B}(x)\! &=&
   N_f\: \{ (1-x) ( -704.67 \: x^3 + 3310.32\: x^2 + 2144.81\: x 
   - 244.68) 
  \nonumber \\ & & \mbox{} 
   + 4490.81\: x^2 L_0 + 42.875\: L_0 - 11.0165\: L_0^3 \} 
\eea
and
\bea
  P^{(2)}_{{\rm PS}, A}(x)\! &=&
   N_f\: \{ (1-x)(-229.497\: L_1 - 722.99\: x^2 + 2678.77 
   - 560.20\: x^{-1})
  \nonumber \\ & & \mbox{}
   + 2008.61\: L_0 + 998.15\: L_0^2 - 3584/27\: x^{-1}L_0 \}
   \:\: + \:\: P^{(2)}_{{\rm PS},2}(x) 
  \nonumber\\
\label{ps1}
  P^{(2)}_{{\rm PS}, B}(x)\! &=&
   N_f\: \{ (1-x) (73.845\: L_1^2 + 305.988\: L_1 + 2063.19\: x 
   - 387.95\: x^{-1}) 
  \nonumber \\ & & \mbox{}
   + 1999.35\: xL_0 - 732.68\: L_0 - 3584/27\: x^{-1}L_0 \}
   \:\: + \:\: P^{(2)}_{{\rm PS},2}(x) 
\eea
with
\bea
\label{ps2}
  P^{(2)}_{{\rm PS},2}(x)\! & = & 
   N_f^2\: \{ (1-x)(-7.282\: L_1 - 38.779\: x^2 + 32.022\: x - 6.252
   + 1.767\: x^{-1} ) \quad
  \nonumber \\ & & \mbox{}
   + 7.453\: L_0^2 \} \:\: .
\eea
The $(\ln x)/x$ term in Eq.~(\ref{ps1}) has been derived in ref.\
\cite{CH94}.

Our new approximations for the off-diagonal singlet splitting 
functions $P^{(2)}_{qg}$ and $P^{(2)}_{gq}$ are given by
\bea
  P^{(2)}_{qg, A}(x)\! &=&
   N_f\: \{ - 31.830\: L_1^3 + 1252.267\: L_1 + 1999.89\: x + 1722.47  
   + 1223.43\: L_0^2
  \nonumber \\ & & \mbox{} 
    - 1334.61\: x^{-1} - 896/3\: x^{-1}L_0 \}
   \:\: + \:\: P^{(2)}_{qg,2}(x) 
  \nonumber\\
\label{qg1}
  P^{(2)}_{qg, B}(x)\! &=&
   N_f\: \{ 19.428\: L_1^4 + 159.833\: L_1^3 + 309.384\: L_1^2 
   + 2631.00\: (1-x) 
  \nonumber \\ & & \mbox{} 
   - 67.25\: L_0^2 - 776.793\: x^{-1} - 896/3\: x^{-1}L_0 \}
   \:\: + \:\: P^{(2)}_{qg,2}(x) 
\eea
with
\bea
\label{qg2}
  P^{(2)}_{qg,2}(x)\! &=&
   N_f^2\: \{ - 0.9085\: L_1^2 - 35.803\: L_1 - 128.023 + 200.929\: 
   (1-x) 
  \nonumber \\ & & \mbox{} 
   + 40.542\: L_0 + 3.284\: x^{-1} \} \:\: ,
\eea
and 
\bea
  P^{(2)}_{gq, A}(x)\! &=&
   13.1212\: L_1^4 + 126.665\: L_1^3 + 308.536\: L_1^2 + 361.21
   - 2113.45\: L_0 
  \nonumber \\ & & \mbox{} - 17.965\: x^{-1}L_0
   \:\: + \:\: N_f\: \{
   2.4427\: L_1^4 + 27.763\: L_1^3 + 80.548\: L_1^2 
  \nonumber \\ & & \mbox{} 
   - 227.135 - 151.04\: L_0^2 + 65.91\: x^{-1}L_0 \} \:\: + \:\:
   P^{(2)}_{gq ,2}(x) 
  \nonumber\\
\label{gq01}
  P^{(2)}_{gq, B}(x)\! &=& \,
   - 4.5108\: L_1^4 - 66.618\: L_1^3 - 231.535\: L_1^2 - 1224.22\:
   (1-x) + 240.08\: L_0^2 \quad
  \nonumber \\ & & \mbox{}
   + 379.60\: x^{-1} (L_0 +4) \: + \:N_f \{
   - 1.4028\: L_1^4 - 11.638\: L_1^3 + 164.963\: L_1 
  \nonumber \\ & & \mbox{} 
   - 1066.78\: (1-x)- 182.08\: L_0^2 + 138.54\: x^{-1} (L_0 +2) \} 
   \:\: + \:\: P^{(2)}_{gq ,2}(x) 
\eea
with
\bea
\label{gq2}
  P^{(2)}_{gq, 2}(x)\! & = &
   N_f^2\: \{ 1.9361\: L_1^2 + 11.178\: L_1 + 11.632 - 15.145\: (1-x) 
   + 3.354\: L_0 
  \nonumber \\ & & \mbox{} 
   - 2.133\: x^{-1} \} \:\: .
\eea
Unlike the case of $P^{(2)}_{qg}$ discussed above, the coefficients 
of the leading small-$x$ terms $(\ln x)/x$ have not been derived for 
$P^{(2)}_{gq}$ up to now. Thus these coefficients in Eq.~(\ref{gq01}) 
have been determined, as before \cite{NV2}, from the available moments.

The $1/[1-x]_+$ soft-gluon contributions to the one- and two-loop
gluon-gluon splitting functions, $P_{gg}^{(0)}$ and $P_{gg}^{(1)}$, are 
related to their quark-quark (non-singlet) counterparts by a factor 
$C_A/C_F$. The same holds for the $N_f^2$ terms at third order
\cite{Gra1,Gra2}. Assuming that this relation holds generally for 
$P_{gg}^{(2)}$, the results given below Eq.~(\ref{nsp}) can be 
employed. In this way we arrive at the approximate expressions
\pagebreak
\bea
  P^{(2)}_{gg, A}(x)\! &=&
   2626.38\: (1-x)_+^{-1} + 4424.168\: \delta (1-x) - 732.715\: L_1^2 
   - 20640.069\: x
  \nonumber \\ & & \mbox{}
   - 15428.58\: (1-x^2) - 15213.60\: L_0^2 + 16700.88\: x^{-1} 
   + 2675.85\: x^{-1}L_0
  \nonumber \\ & & \mbox{} \hspace{-13mm}
   + N_f\: \{ - 415.71\: (1-x)_+^{-1} - 548.569\: \delta (1-x)
   - 425.708\: L_1 + 914.548\: x^2 
  \nonumber \\ & & \mbox{} 
   - 1122.86 - 444.21\: L_0^2 + 376.98\: x^{-1} 
   + 157.18\: x^{-1}L_0 \} \:\: + \:\: P^{(2)}_{gg,2}(x) 
  \nonumber\\
\label{gg01}
  P^{(2)}_{gg, B}(x)\! &=&
   2678.22\: (1-x)_+^{-1} + 4590.570\: \delta (1-x)
   + 3748.934\: L_1 + 60879.62\: x
  \nonumber \\ & & \mbox{}
   - 35974.45\: (1+x^2) + 2002.96\: L_0^2 + 9762.09\: x^{-1}
   + 2675.85\: x^{-1}L_0
  \nonumber \\ & & \mbox{} \hspace{-13mm}
   + N_f\: \{
   - 412.00\: (1-x)_+^{-1} - 534.951\: \delta (1-x)
   + 62.630\: L_1^2 + 801.90 
  \nonumber \\ & & \mbox{} 
   + 1891.40\: L_0 + 813.78\: L_0^2 + 1.360\: x^{-1} 
   + 157.18\: x^{-1}L_0 \} \:\: + \:\: P^{(2)}_{gg ,2}(x) 
   \quad
\eea
with
\bea
\label{gg2}
  P^{(2)}_{{gg}, 2}(x)\! & = &
   N_f^2\: \{ -16/9\: (1-x)_+^{-1} + 6.4882\: \delta (1-x)
   + 37.6417\: x^2 - 72.926\: x 
  \nonumber \\ & & \mbox{} 
   + 32.349 - 0.991\: L_0^2 + 2.818\: x^{-1} \} \:\: .
\eea
The $(\ln x)/x$ terms in Eq.~(\ref{gg01}) have been determined in ref.\
\cite{FL98} in a scheme equivalent to the DIS scheme up to NNLO.
The transformation to \MSb\ can be found in ref.~\cite{NV2}.

The uncertainty bands for the three-loop singlet splitting functions 
resulting from Eqs.~(\ref{ps1})--(\ref{gg2}) are displayed in Fig.~3
for $N_f = 4$. As in the non-singlet cases considered above, our new 
parametrizations considerably improve on the previous uncertainties. 
While the small-$x$ behaviour of our approximations obviously depends
on the results of refs.~\cite{CH94,FL98}, it is worthwhile to note that
reducing the $N_f^0$ coefficient of $(\ln x)/x$ in $P^{(2)}_{gg}$ by a
factor of two does not lead to approximations outside the error band
in Fig.~3.

In Fig.~4 we finally illustrate the impact of the NNLO terms on the 
scale derivatives~(\ref{sgev}) of the singlet quark and gluon 
densities. As in ref.~\cite{NV2} the initial conditions are chosen~as
\bea
\label{pinp}
 x\Sigma (x,\mu_{f}^2) &\! =\! & 0.6\, x^{-0.3}\, (1-x)^{3.5}
                                   (1 + 5\, x^{0.8})  
 \nonumber \\
 xg (x,\mu_{f}^2) &\! =\! & 1.0\, x^{-0.37} (1-x)^{5} 
\eea
and
\beq
\label{ainp}
  \alpha_s (\mu_r^2 = \mu_{f}^2) \: = \: 0.2 \:\: ,
\eeq
corresponding to $\mu_f^2 \simeq 30 \mbox{ GeV}^2$. Under these
conditions the residual uncertainties of the three-loop contributions 
amount to about $\pm 2\%$ or less down to $x \simeq 10^{-4}$, even if 
the bands in Fig.~4 were increased by 50\% in order to account for any 
possible underestimate of the errors. At lower scales the flatter 
small-$x$ shapes of the quark and gluon densities, together with the 
larger $\alpha_s$, lead to larger uncertainties at small $x$. At 
$x= 10^{-4}$ and $\mu_{f}^2 \simeq 3\mbox{ GeV}^2$, for example, they 
reach about $\pm 4\%$ and $\pm 3\%$ for the singlet quark and gluon 
derivatives, respectively, for standard NLO distributions like CTEQ4M
\cite{CTEQ}. These numbers represent improvements by about a factor of
three on our previous results \cite{NV2}. 
Thus our present approximations, based on the results of ref.~\cite{RV},
facilitate a reliable NNLO evolution of unpolarized parton 
densities down to, at least, $\mu_f^2 \,\gsim\, 10 \mbox{ GeV}^2$ and 
$x \,\gsim\, 10^{-4}$.
 
\noindent
{\sc Fortran} subroutines of the above approximations of the
three-loop splitting functions can be obtained via email to 
neerven@lorentz.leidenuniv.nl or avogt@lorentz.leidenuniv.nl.
%
%
\newpage
\section*{Acknowledgements}
%
%
We are grateful to J. Vermaseren for communicating the results of ref.\
\cite{RV} to us prior to publication. We also thank V. Braun for 
pointing ref.~\cite{Ko89} out to us.
This work has been supported by the European Community TMR research
network `QCD and the Deep Structure of Elementary Particles' under 
contract No.~FMRX--CT98--0194.
%
%

%
%
\newpage
\vspace*{\fill}
\centerline{\epsfig{file=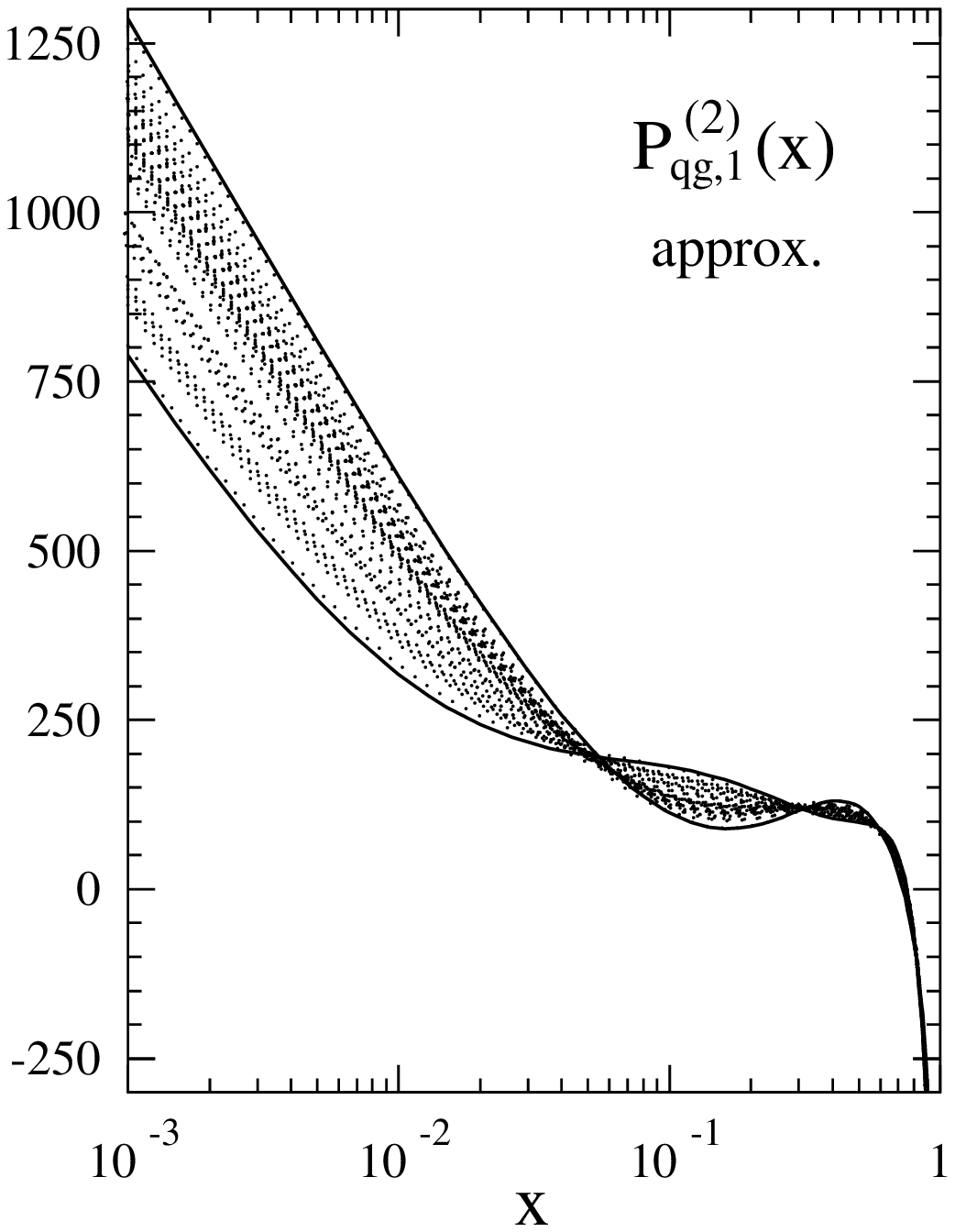,width=12cm,angle=0}}
\vspace{2mm}
\noindent
{\bf Figure 1:} Approximations of the $N_f^1$ part $P_{qg,1}^{(2)}$ of 
 the three-loop splitting function $P_{qg}^{(2)}(x)$, as obtained from 
 the six lowest even-integer moments \cite{RV,spfm2} together with the 
 leading small-$x$ term of ref.~\cite{CH94}. The full curves represent 
 those functions selected for Eq.~(\ref{qg1}).
\vspace*{\fill}

\newpage
\vspace*{\fill}
\centerline{\epsfig{file=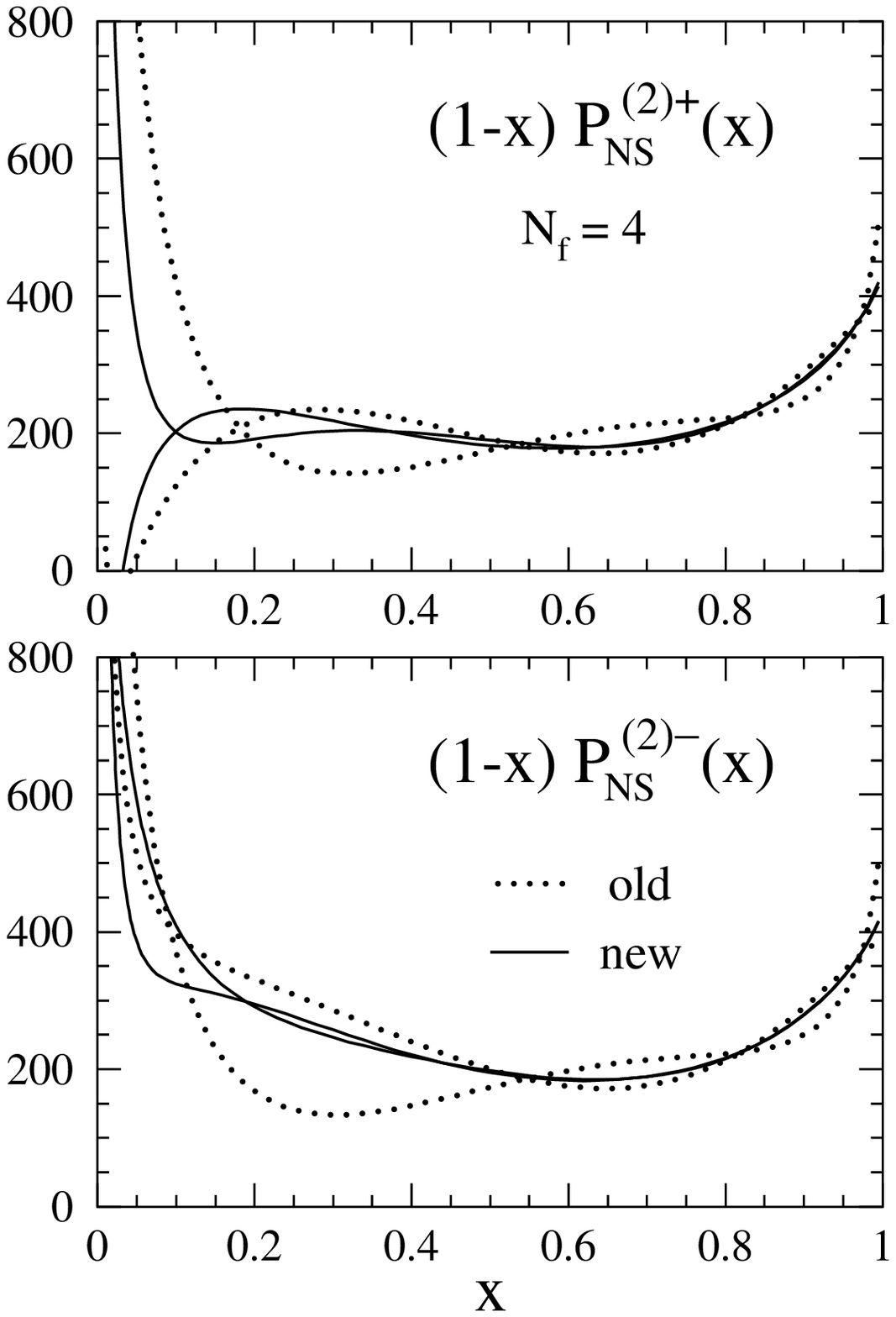,width=12cm,angle=0}}
\vspace{2mm}
\noindent
{\bf Figure 2:} Top: Our new approximations of $P_{\rm NS}^{(2)+}(x)$ 
 for $N_f=4$, as obtained from Eq.~(\ref{nsp}) together with Eq.~(4.13) 
 of ref.~\cite{NV1}. The dotted curves represent our previous 
 parametrizations \cite{NV1}. 
 Bottom: The same for $P_{\rm NS}^{(2)-}(x)$ using Eq.~(\ref{nsm}).
\vspace*{\fill}

\newpage
\vspace*{\fill}
\centerline{\epsfig{file=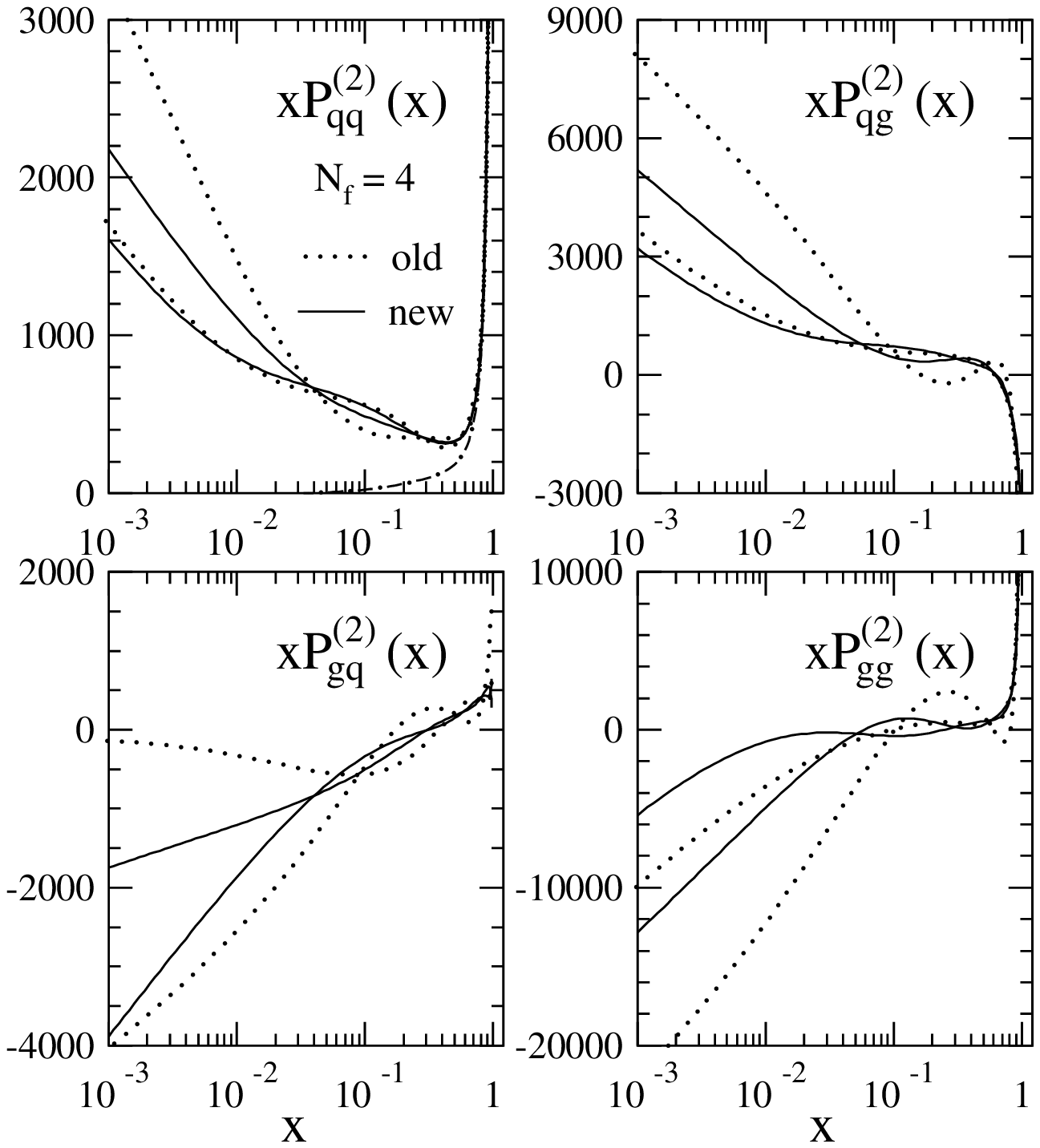,width=13.7cm,angle=0}}
\vspace{2mm}
\noindent
{\bf Figure 3:} Our new approximations of the singlet splitting 
 functions $P_{ij}^{(2)}(x)$ for $N_f=4$. $P_{qq}^{(2)}$ is obtained by 
 adding $P_{\rm NS}^{(2)+}(x)$ of Fig.~2 (separately shown by the 
 dash-dotted curve) and $P_{\rm PS}^{(2)}(x)$ of Eqs.~(\ref{ps1}) and 
 (\ref{ps2}). Our previous parametrizations \cite{NV2} are also shown.
\vspace*{\fill}

\newpage
\vspace*{\fill}
\centerline{\epsfig{file=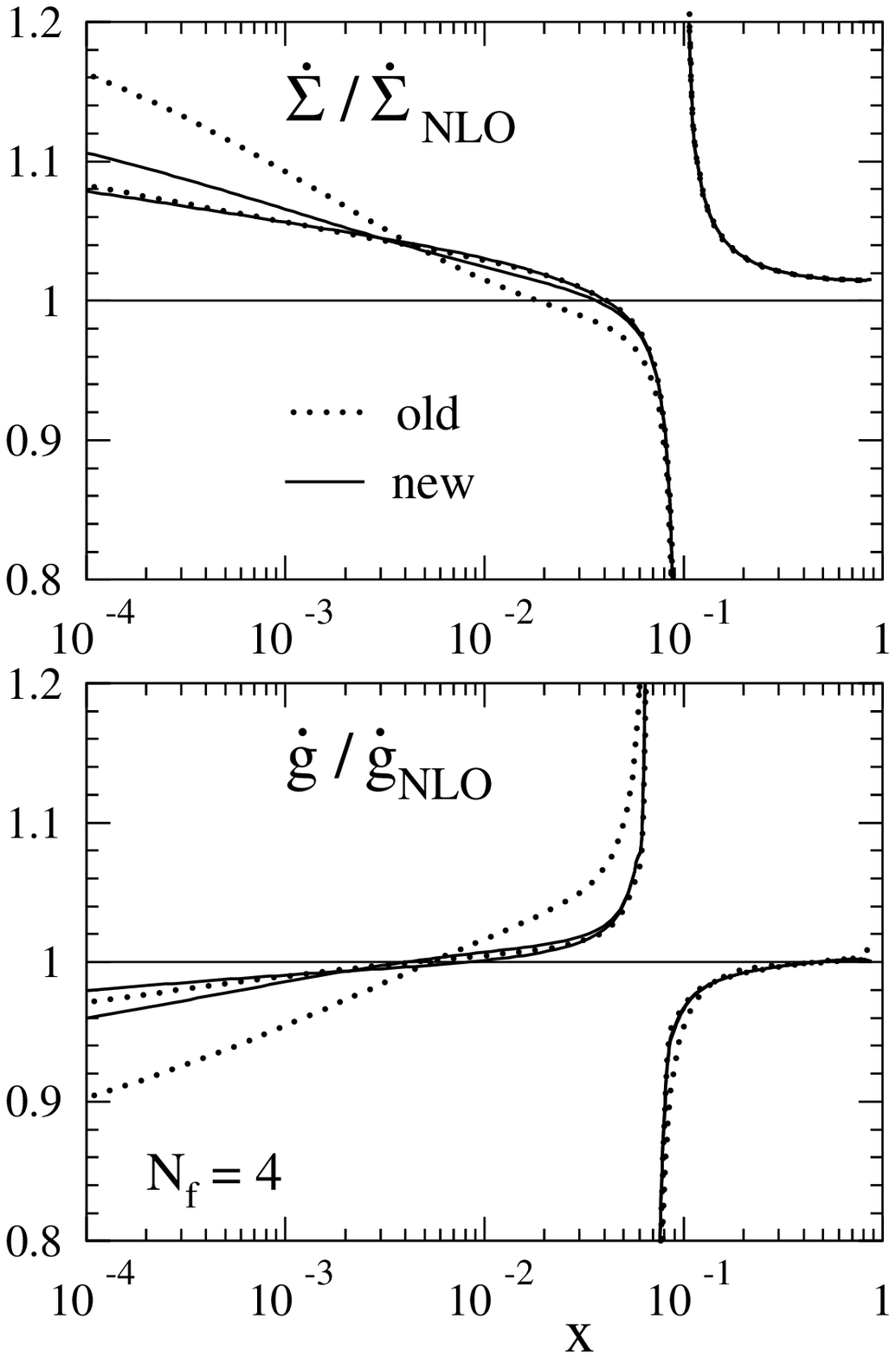,width=12cm,angle=0}}
\vspace{2mm}
\noindent
{\bf Figure 4:} The size and remaining uncertainties of the NNLO 
 corrections for the scale derivatives, $\dot{\Sigma} \equiv d \Sigma / 
 d\ln \mu_f^2$ and $\dot{g} \equiv dg / d\ln \mu_f^2$, of the singlet 
 quark and gluon densities at $\mu_f^2= \mu_r^2\simeq 30 \mbox{ GeV}^2$ 
 ($\alpha_s = 0.2$). The input densities are specified in 
 Eq.~(\ref{pinp}).
\vspace*{\fill}
%
%
\end{document}